# The first eclipsing B-shell + sdO binary discovered in HD 328058

Norbert Hauck

Bundesdeutsche Arbeitsgemeinschaft für Veränderliche Sterne e.V. (BAV), Munsterdamm 90, 12169 Berlin, Germany; hnhauck@yahoo.com

**Abstract:** *An eclipsing binary with an orbital period of 62 days has been discovered in HD 328058 and studied by UBVIcHα photometry. Combining the results of light curve modelling with existing stellar models and known B3-shell spectral type of the primary star then allowed a reliable estimation of the mass, radius and effective temperature of both components: about 6.9 Msun, 5.2 Rsun and 17600 K for the primary star, and about 0.78 Msun, 1.20 Rsun and 56000 K for the hot subdwarf companion of type sdO. Obviously, this binary has been created by mass transfer of the envelope of the sdO progenitor to our primary star, now being a rapidly rotating shell star. Its spheroidal distortion and its decretion disk can be studied here unusually well.*

HD 328058 being located in the constellation of Norma is described as a single star in the SIMBAD database. The only literature reference therein (Nesterov et al. [1]) gives a spectral type B3. Bidelman et al. [2] have attributed spectral type B shell and mentioned spectral variability. From GAIA-DR2 we get a preliminary distance of 951 (908 - 999) pc. Data mining and processing of old photometry data from ASAS-3 has now revealed that this stellar object is indeed an eclipsing binary with an orbital period of 62 days (see Fig. 1).

A remote controlled 0.5m-reflector telescope in Siding Spring, Australia, has been used to obtain more precise and complete photometric data in passbands UBVIcHα. HD 328064 has been the main comparative star. The primary minimum of the light curve is located between the phases ≈ 0.973 and 0.027, and is showing a larger outer part obviously created by absorption of light of the smaller secondary component by the decretion disk seen edge-on (resp. 'shell') surrounding the equator of our B-type primary star. The inner resp. central part of the minimum between phase 0.9943 and 0.0057 is caused by the stellar eclipse, when the companion star is covered by the primary star (see Fig. 1). The shallow secondary minimum has not been found in the data of ASAS-3 or ASAS-SN, however, finally been detected at phase 0.50 in our new data. In the model created with the *Binary Maker 3* (BM 3) software (Bradstreet & Steelman, 2004) it has a depth of 16, 21, 27 and 30 mmag in passbands UBVIc, respectively. This has been confirmed by our photometry in Ic. The somewhat smaller than predicted depths measured at shorter wavelengths λ might be caused by a larger limb darkening effect (λ) of the distorted shell star, compared to the standard model used in the BM 3.

Our new photometric data have been used for light curve modelling of the central star eclipses with the *Binary Maker 3* software. The best fit of the primary minimum achieved a σ-Fit of 1.5 mmag for a circular orbit (see Fig. 2). The partial eclipse is deep (≈ 76 % of totality) and allows a detailed study of the rotational distortion of the primary star to an oblate spheroid (see to-scale Fig. 3.1 and 3.2). The ratio of its equatorial to polar radius was found to be 5.58 R☉ / 4.34 R☉, i.e. ≈ 1.29 at a fitted super-synchronous rotation factor of 81.1.

The permanent absorption of primary star's light by half of its decretion disk, and the only temporary absorption of the light of the secondary star by the entire disk diameter at primary minimum require separate modelling of both minima. The entire disk causes a significant loss of light of the secondary star of about 59% in passband U, 53% in B, 51% in V, 58% in Ic, and 72% in H alpha, which is regarded as an indication of a decretion disk consisting of relatively dense gas. Our eclipse geometry requires a disk having a rim width of at least primary star's polar diameter, and therefore having a remarkably large opening angle (starting at the equator) in a cross-sectional view.

Spectral type B3 (from ref. [1]) is equivalent to a $T_{eff}$ of about 17000 K in our equatorial view. Comparative modelling with the BM3 gives a mean $T_{eff}$ of 17600 K over the full surface area, i.e. including the hotter polar regions of the rapidly rotating star. In a stellar model with no rotation the theoretical mean $T_{eff}$ is higher, i.e. about 19200 K.

Secondary star's $T_{eff}$ has been investigated by modelling the light curves of the central primary minimum. The first results showed clear wavelength (λ) dependence, i.e. a $T_{eff}$ of 33, 37, 42 and 45 kK in the passbands UBVIc, respectively. This effect is regarded as reddening of the secondary star by a thick and dusty accretion disk, possibly puffed up by its intense radiation pressure, and has been incorporated in the model. Therefore, the extinction A(λ) has been compensated by reducing the size of this star (in an artificial total eclipse), and strictly in line with the extinction law and coefficients of Cardelli et al. [3]. Thereby the $T_{eff}$ has been successfully converged to a single value (± 250 K) in all passbands UBVIc simultaneously. For an adopted maximum $R_V$ (= $A_V$ / $E_{B-V}$) of 6, the highest known value measured so far, an acceptable fit has been achieved only for $T_{eff}$ ≥ 51 kK, which hence should be our lower limit for this parameter. At higher $T_{eff}$ a slightly lower $R_V$ delivered better fits: e.g. at a $T_{eff}$ of 56 kK an $R_V$ of 5.7, and at 60 kK an $R_V$ of 5.5. Secondary star's extreme $T_{eff}$ together with its solar-like mass and radius (see here below) are indications of a hot subdwarf star of type sdO, i.e. a helium star containing a carbon/oxygen core and only a very thin hydrogen coating (for a review see Heber [4]).

The absolute dimensions of the shell star have been derived from the stellar models of Ekström et al. [5] for solar metallicity (Z = 0.014) and no rotation. Therefore, the relative parameters obtained from light curve modelling have been transformed into their corresponding values for a non-rotating star with help of equations and diagrams given in Ekström et al. [6]. Our radius / separation ratio R/a then fits to a single position in the middle of the main sequence band of their Hertzsprung-Russell diagram (HRD) for our $T_{eff}$ (and calculated separation a from Kepler's third law). Thereby a mass of 6.87 ± 0.52 M☉ has been obtained for the shell star. The relatively large radius of the sdO of 1.20 (+0.21/-0.11) R☉ has then been derived from the size of the primary star. The mass of 0.78 ± 0.05 M☉ for the sdO has been taken from evolutionary tracks for helium stars at $Z_{solar}$ given in Fig. 3.4 of Heusgen [7]. Masses below 0.73 M☉ have been excluded, since they never (or too late) achieve our radius of 1.20 R☉. Masses above 0.83 M☉ evolve to larger radii in less than 30 Myr, i.e. before the minimum age of our primary star. Moreover, for our mass and radius range, these evolutionary tracks limit $T_{eff}$ of the sdO to about ≤ 61 kK.

In the most likely scenario our binary has been created by Roche-lobe overflow of the hydrogen-rich envelope of the former mass donor (loser) of an initial mass of (almost) 5 Mo to its mass accreting companion (gainer) of initially ≈ 3 Mo. From the stellar models of Ekström et al. [5] it can be derived, that thereafter it took ≈ 31 Myr plus a few more for our rapidly rotating shell star to reach its present post-Algol evolutionary stage. This fits quite well to our sdO-helium remnant of the donor star, which according to an empirical equation of Z. Han (cited by Eggleton [8]) has an expected lifetime of about 30 Myr in the helium main sequence. Subsequently, in the remaining timespan difference, our sdO has evolved to its present stage in the HRD. According to the HRD given in Fig. 3.4 of Heusgen [7] our hot subdwarf, after a previous expansion and cooling step, is now contracting at nearly constant top luminosity until a remarkable $T_{eff}$ of ≥ 200000 K is reached. Finally, it will follow the white dwarf (WD) cooling track.

To date there have been discovered five similar binaries of the type Be + sdO by analysis of far-ultraviolet spectra from space-based observations (see e.g. table 3 in Schootemeijer et al. [9]). Wang et al. [10] have added recently twelve candidates for this type of binaries containing primary stars of spectral type B0 to B3. Although no emission lines have been reported yet, our B-shell + sdO binary should belong to this group, since a Be star seen equator-on is a shell star (see Rivinius et al. [11]). Moreover, the derived equatorial rotational velocity of ≈ 367 km/s is equivalent here to a fraction W = 0.76 of the critical value, and fitting well to a Be-type star (see Fig. 9 of Rivinius et al. [12]). HD 328058 is the first eclipsing binary of this type allowing detailed studies of the decretion disk, when the light of the sdO passes through. Our sdO's $T_{eff}$ of about 56 kK is slightly above the known range of 42 to 53 kK for the sdO components in these binaries. After V658 Car (HD 92406), obviously in the earliest post-Algol evolutionary stage ever seen (see Hauck [13]), HD 328058 is a new clear finding of a shell star in an eclipsing binary. However, it does contain more massive components: the low-mass He-WD precursor of V658 Car here being replaced by a helium-burning C/O-WD precursor.

The B–V colour excess $E_{B-V}$ of ≈ 0.50 for our binary indicates a significant reddening corresponding to an extinction $A_V$ of ≈ 1.57 mag in our line of sight, which might be the effect of an interstellar dust cloud in front of HD 328058 and his stellar neighbour HD 328059 (in 1.5' angular distance, and showing an extinction $A_V$ of 1.19 mag). A circumbinary disk being formed by mass loss of the system during former non-conservative mass transfer might be a further explanation. The calculated distance of about 1050 pc for HD 328058 is about 10% above the mean value of 951 pc of the second data release of the GAIA mission, which appears satisfactory for the time being, given their ranges of error found for hot and cool stars having similar parallaxes (see Fig. 1 of Stassun et al. [14]). The results are listed in tables 1 and 2. The error margins are based on 5% uncertainty in primary star's $T_{eff}$ and on an adopted solar composition.

Further studies of HD 328058 appear interesting. This eclipsing binary system should allow double-lined spectroscopy for getting precise masses. The $T_{eff}$ parameters could be determined with higher accuracy by modern spectroscopic means and stellar atmospheric models. The rapidly evolving primary star can be used as a scale for the age determination of the sdO component. Hence existing stellar models can be tested here unusually well.

**Table 1: Parameters of binary system HD 328058**

| | | |
|---|---|---|
| Epoch [HJD] | 2452494.305(3) | mid primary minimum |
| Period P [days] | 62.469(2) | from ASAS + new data |
| Maximum light V/B [mag] | 9.80 / 10.03 | from ASAS-3 [15] / TYCHO |
| Minimum light in V [mag] | 10.125 / 9.825 | primary / secondary eclipse |
| Eclipse duration [hours] | 81 / 17.1 | disk / stellar eclipse |
| Orbital inclination i [deg] | 88.33 | (+0.02/-0.05) |
| Orbital radius a [R☉] | 130.4 ± 2.9 | for R☉ = 696342 km; circular orbit |
| Distance [pc] | 1050 | (+150/-112); for $A_v$ = 1.57 mag |

**Table 2: Parameters of components of HD 328058**

| Parameter | Primary star | Secondary star | Decretion disk |
|---|---|---|---|
| Spectral type | B3 shell | O adopted | |
| Temperature $T_{eff\ mean}$ [K] | 17600 ± 900 | 56000 ± 5000 | |
| Radius (R pole) [R☉] | 4.34 ± 0.10 | | |
| Radius (R equator) [R☉] | 5.58 ± 0.13 | | |
| Radius (R mean) [R☉] | 5.17 ± 0.12 | 1.20 ± 0.16 | 21.1 ± 0.8 |
| Luminosity (bol.) [log L☉] | 3.35 ± 0.11 | 4.10 ± 0.18 | |
| Brightness (abs.) [VMag] | - 1.50 dimmed | - 0.55 | |
| V-light fraction at Max. | 0.706 dimmed | 0.294 | 0 adopted |
| Mass [M☉] | 6.87 ± 0.52 | 0.78 ± 0.05 | |

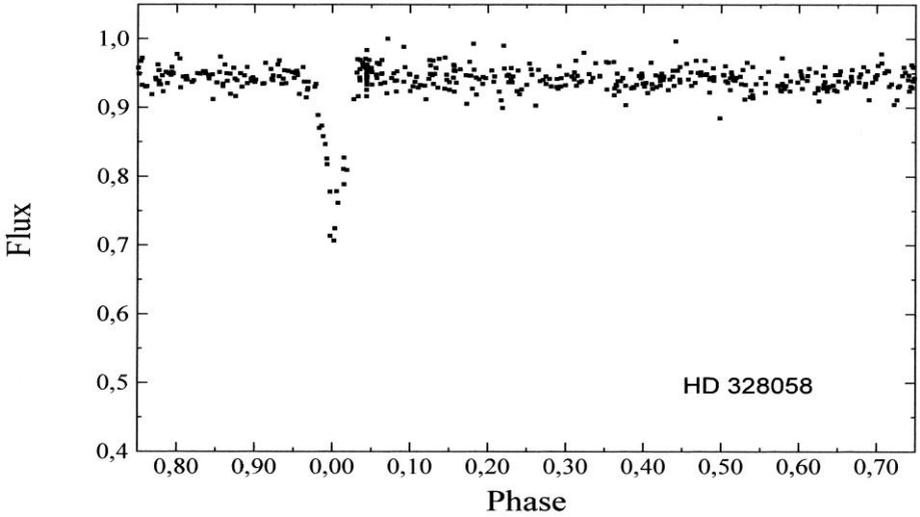

**Fig. 1:** Phase – VFlux plot with primary minimum for P=62.469 days from ASAS-3 data

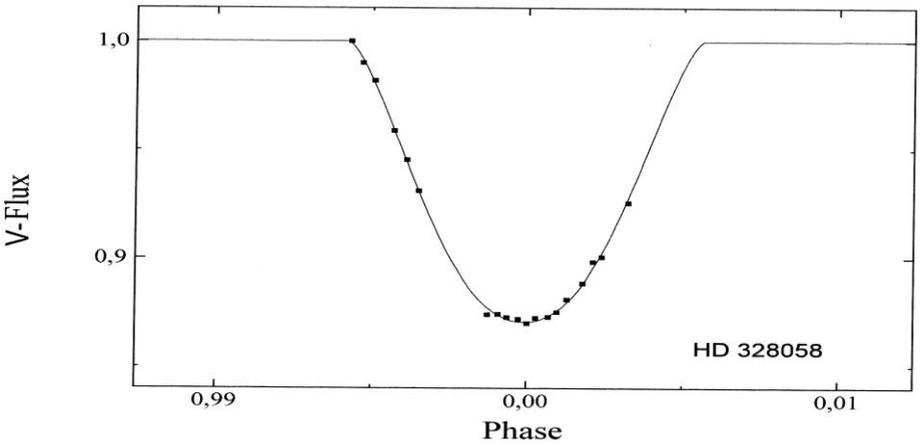

**Fig. 2:** Computed light curve of the central stellar eclipse for our new, binned V-data from JD 2458241 to 2458304

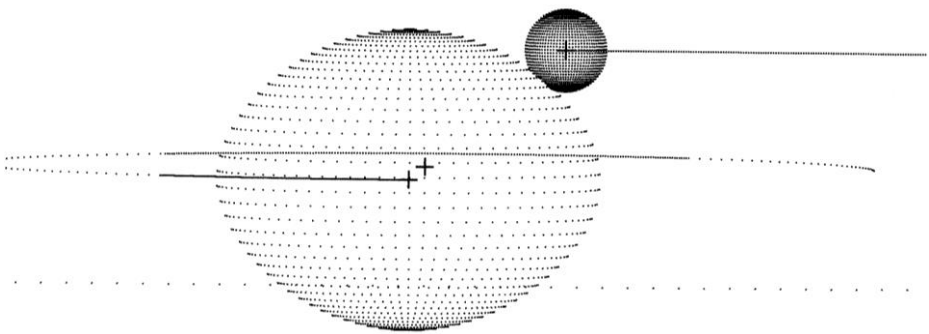

**Fig. 3.1:** HD 328058 shortly after first contact of stellar primary eclipse (phase 0.9943).

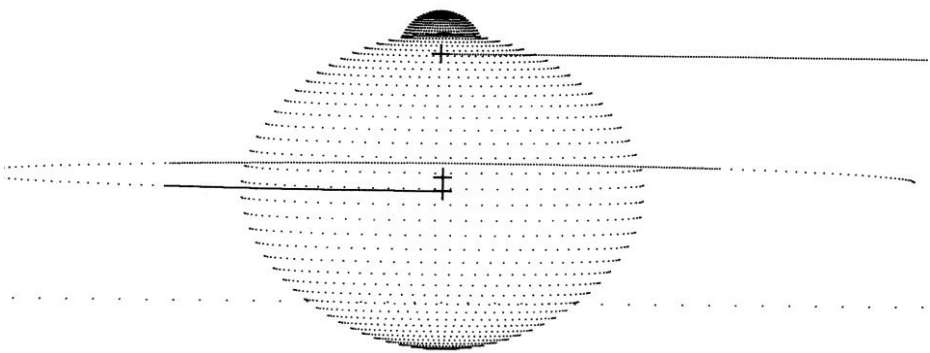

**Fig. 3.2:** HD 328058 at mid stellar primary minimum (phase 0). To-scale figures of the rapidly rotating B-shell star and its sdO companion in our equatorial view. Gravity centres (+) and their orbital movements are also indicated.

**Acknowledgements:**
This research has made use of the Simbad and VizieR databases operated at the Centre de Données astronomiques de Strasbourg, France, http://cdsarc.u-strasbg.fr/ and the All Sky Automated Survey ASAS database, http://www.astrouw.edu.pl/asas